\documentclass[a4paper,10pt]{article}

\usepackage{graphicx}
\usepackage{amsfonts}
\usepackage{amsthm}
\usepackage{amsmath}
\usepackage{amssymb}

\usepackage[top=1in, bottom=1in, left=1in, right=1in]{geometry}

\usepackage{graphicx}
\usepackage{epstopdf}
\usepackage[center,bf]{caption}

\usepackage{authblk}

\begin{document}


\title{Do neutrons publish? A neutron publication survey 2005-2015}

\author
       {T. Gutberlet$^1$, D. Tunger$^2$, P. Zeitler$^2$,
       T. Br\"uckel $^1$\\
       $^1$J\"ulich Centre for Neutron Science, Forschungszentrum J\"ulich GmbH,   
			 J\"ulich, Germany\\
       $^2$Zentralbibliothek, Forschungszentrum J\"ulich GmbH, J\"ulich, Germany\\
       \textit{Corresponding author: t.gutberlet@fz-juelich.de}\\ 
			 }

\maketitle

\begin{abstract}
\noindent
Publication in scientific journals is the main product of scientific research. The amount of papers published, their placement in high impact journals, and their citations are used as a measure of the productivity of individual scientists, institutes or fields of science. To give a profound basis on the publication record and the quality of the publication efforts in neutron scattering, a survey has been done following the approach to use bibliographic databases. Questions to be addressed by this survey are: Is the productivity of research with neutrons changing over the years? Which is the geographic distribution in this field of research?  Which ones are leading facilities? Is the quality of publications changing? The main results found are presented.
\end{abstract}


\section{Introduction}

Publication in scientific journals is the main product of scientific research. In peer reviewed journals the manuscript is refereed prior to publication by experts in the corresponding field of science. Published papers of interest are cited in other publications by the scientists, which distributes a publication within the scientific community. Journals which have a significant reputation in a scientific area are preferred by authors as they may help to reach out to as many of interested scientists in the particular field. This helps to level the impact of a paper and in response of the scientific journal.  The amount of papers published, their placement in high impact journals, and their citations are used as a measure of the productivity of individual scientists, institutes or fields of science. 

The European Neutron Scattering association estimates about 1900 peer reviewed articles each year by European researchers \cite{1,2}. A number of about 2300 peer reviewed articles per year were given in a report by the ILL \cite{3}. These numbers might be an indication of the impact neutron scattering seems to have in Europe with a user base of about 8000 researchers \cite{1}. Categorizing publications according to the method used, one has to take the availability of the method into account. For neutron scattering the access is limited due to the low number of neutron sources, the number of neutron instruments installed and the necessary time for an experiment. Clearly this capacity restricts the possible publication output. Another factor regarding impact is the quality of the published papers in journals with high impact factors as Nature, Science etc. About 23\% of the papers in the report by ILL were related to papers published in such high impact journals \cite{2}. 

There is an ongoing debate on the relevance of publication numbers, impact factors and related indicators as a measure for the quality and impact in science \cite{4,5}. It is clear that simple indicators as the pure number of publications are not sufficient to describe the activity and the impact of a particular field or of a scientific method. Categorizing publications according to the impact factor of journals may be wrong leading as a high impact factor does not automatically lead to a high citation rate. On the other hand relating the publication to its citations only works on longer time scales and may vary with scientific interest or "hot topics" in vogue. Hence, a publication in a low impact journal may outperform over time a paper published in a high impact journal in number of citations. 

To give a profound basis on the publication record and the quality of the publication efforts in neutron scattering, a survey has been done following the approach to use bibliographic databases \cite{6}. The study was also motivated by the changing neutron landscape with older facilities being phased out, new facilities starting up, new communities developing in different world regions, but also new methods such as e.g. high resolution electron microscopy developing. 
The survey covered the period 2005-2015 with a world-wide search on publications listed in the "Web of Science" \cite{7} employing neutron scattering. Questions to be addressed by this survey are: Is the productivity of research with neutrons changing over the years? Which is the geographic distribution in this field of research?  Which ones are leading facilities? Is the quality of publications changing? The main results found are presented.

\section{Method}
\noindent
The survey was performed by the Zentralbibliothek of Forschungszentrum J\"ulich using the Web of Science. The strategy of the survey was based on a basic request with a list of 44 keywords (see Supplement). This gave 64800 hits in Web of Science. This data based was further refined on the basis of 274 Web of Science subject categories of which 81 were included in the search and 11 checked as relevant on the journal level. This reduced the number of hits to 50800. Final refinement by exclusion of journals non relevant for publication of neutron scattering results led to a remaining number of 49769 publications and 658033 related citations in Web of Science in the period of 2005 to 2015. The obtained number of publications and citations were analysed according to the affiliation of all authors given by country and town of the affiliated author institute. A publication with e.g. five authors, three affiliations in three towns and two countries is then listed in each town and in each country once. This analysis has been broken down on world regions (see Appendix). In addition individual countries or towns have been analysed. Certain towns have been related to neutron facilities located to relate the publication record to individual neutron facilities.

\section{Results}

\begin{figure}[tbp]
\centering
\includegraphics[width=0.4\linewidth]{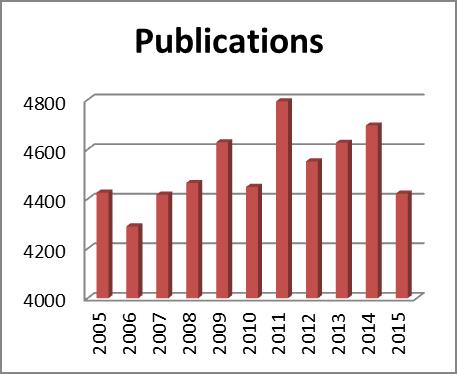}
\caption{Publications per year world wide in neutron scattering.}
\label{Fig1}
\end{figure}

The survey resulted in 49769 individual publications in the period 2005 to 2015. The average publication record per year is 4524, varying between 4290 and 4794 in the survey period (Fig. \ref{Fig1}). A number of 658033 citations was recorded for these publications in the survey period. The average citation record per year was 59821, which gives an average number of citations per publication per year of 13.2 (see Table 1). As the citation cumulates with time, older publications have a higher citation number as younger publications. This is reflected in the raising citation per publication number (CPP) reaching a value of about 20 after 10 years on average. If the citations are normalized to the relative time period, average citations per year are 9986, ranging from about 7516 (citations in 2015) to around 11000 (in years 2009 to 2012).

\begin{figure}[tbp]
\centering
\includegraphics[width=0.4\linewidth]{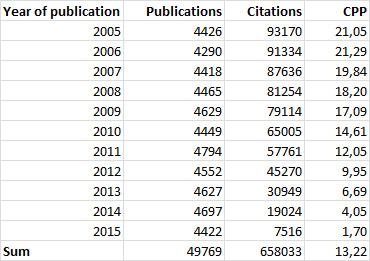}\\

\textbf{Table 1:} Number of publications and citations per year and the average citation per publication (CPP) world wide in neutron scattering.
\end{figure}

\subsection{Publications by world regions} Counting the overall publications in this period by world regions gives a total of 81307 publications. Taken into account the total number of publications of 49769, each publication is attributed to 1,63 countries by the author affiliations. This is nearly twice the number of individual publications due to trans-national multi-author publications counting for each country separate. It demonstrates the good international cooperation characteristic for research with neutrons.

\begin{figure}[tbp]
\centering
\includegraphics[width=0.4\linewidth]{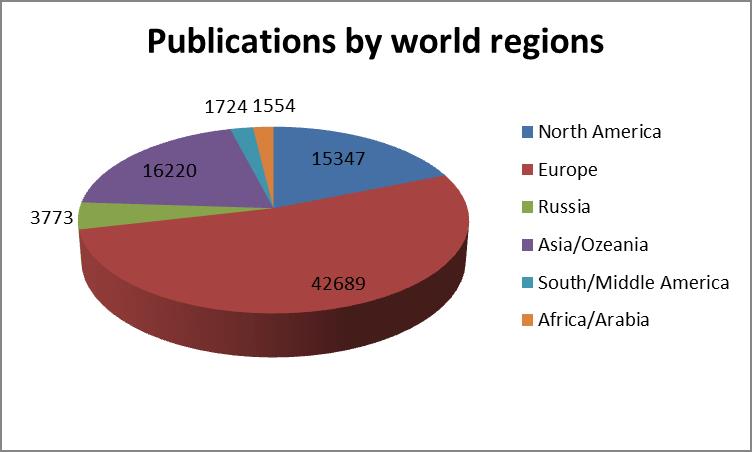}
\caption{Total publications by world regions in neutron scattering between 2005-2015.}
\label{Fig2}
\end{figure}

\begin{figure}[tbp]
\centering
\includegraphics[width=0.4\linewidth]{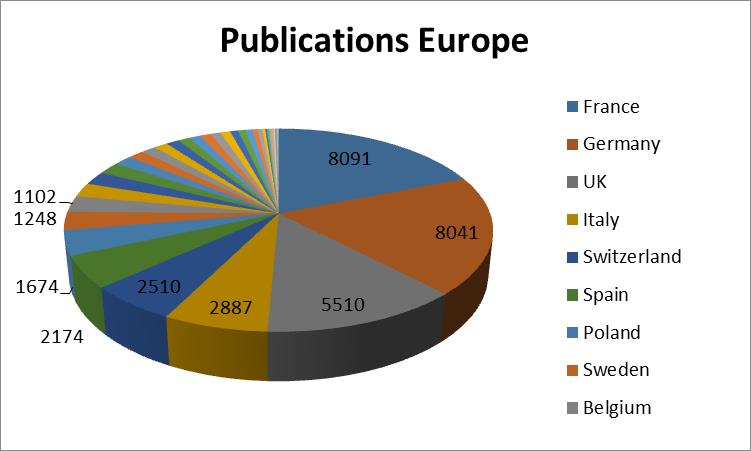}
\caption{Distribution of publications in Europe by country in neutron scattering between 2005-2015.}
\label{Fig3}
\end{figure}

The majority of these counted publications are from Europe (52\%), followed by the US (19\%) and Asia/Oceania (20\%) (Fig. \ref{Fig2}). Within Europe the strongest countries are France (8091), Germany (8041) and the UK (5510) (Fig. \ref{Fig3}). Note that papers with co-authors from the ILL as European facility significantly increase the number of publications for France.

Similar numbers have been given in the report by the ILL looking at the period between 2000 and 2009, also the numbers for the UK were higher with about 7000 \cite{3}. Within the survey period the annual number of publications increased from about 7000 (2005: 6972) to about 8000 (2014: 8099). This increase was different in the different world regions (see below).

\begin{figure}[tbp]
\centering
\includegraphics[width=0.5\linewidth]{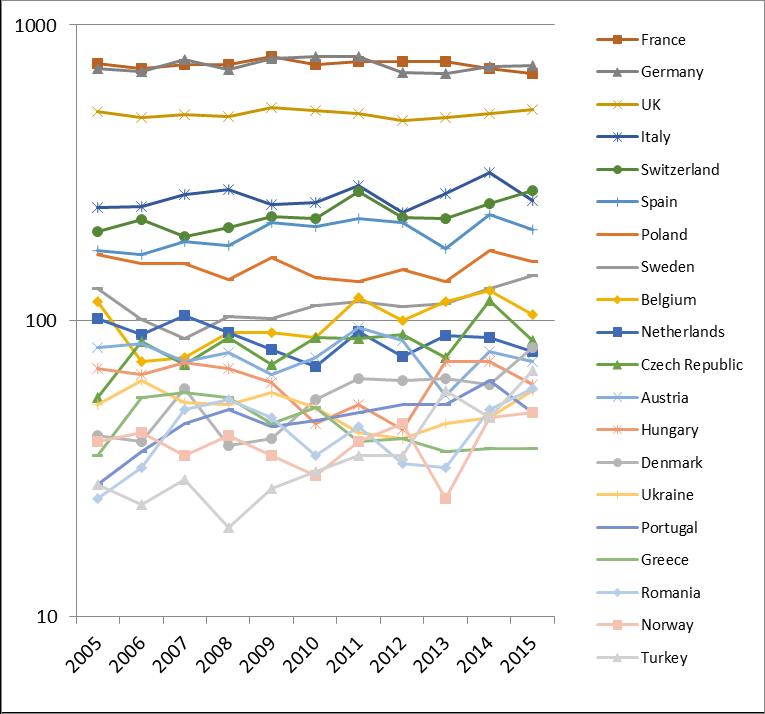}
\caption{Publications in Europe per year in neutron scattering for each country.}
\label{Fig4}
\end{figure}

The average publication number in Europe showed a slight increase from about 3700 at start to about 4000 now. In the period of the survey the publications stay constant for most of the top countries as e.g. for Germany (Fig. \ref{Fig4}). Increases are seen in countries as Switzerland, Sweden or Portugal (Fig. \ref{Fig4}). 

If one relates the number of publications to the size of the country population, Switzerland is the most productive country with 3290 residents/publication, followed by France (8280), Germany (10130) and the UK (11800).

The publication record of the US and Canada show a stable number of about 1400 publications per year in the survey period. A slightly increased number is found between 2007 and2013 with a high in 2013 with 1507 publications. This increase might reflect the establishment of the user operation at the Spallation Neutron Source SNS in Oak Ridge, TN.

\begin{figure}[tbp]
\centering
\includegraphics[width=0.4\linewidth]{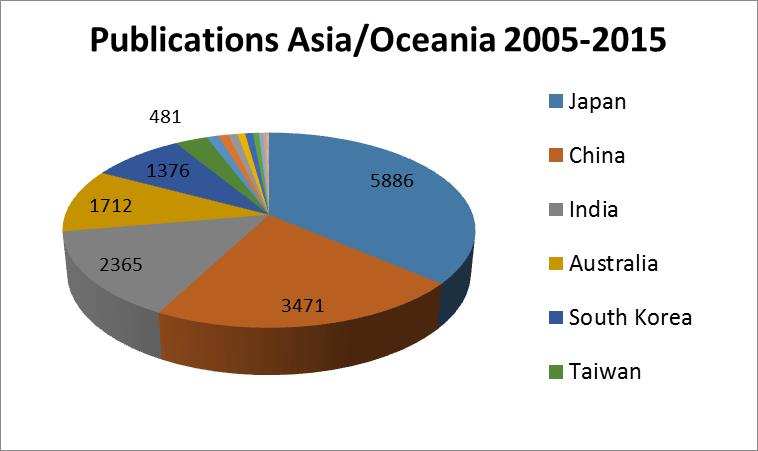}
\caption{Publications in Asia/Oceania per year in neutron scattering between 2005-2015.}
\label{Fig5}
\end{figure}

Asia/Oceania has a contribution of 20\% to the world neutron papers (16220 publications), dominated by Japan (36\%), China (21\%) and India (15\%) (Fig. \ref{Fig5}). There is a strong increase in the publication number from 1209 publications in 2005 to 1647 publications in 2015. While the number of publications in Japan seems to decrease from 565 (2005) to 402 (2015) the numbers are increasing strongly in China in the same period (2005: 209, 2015: 498), so today China is outperforming Japan in this area regarding number of publications. A similar strong increase is found in India (2005: 136, 2015: 265) and Australia (2005: 116, 2015: 194) (see Table 2). 

\begin{figure}[tbp]
\centering
\includegraphics[width=0.9\linewidth]{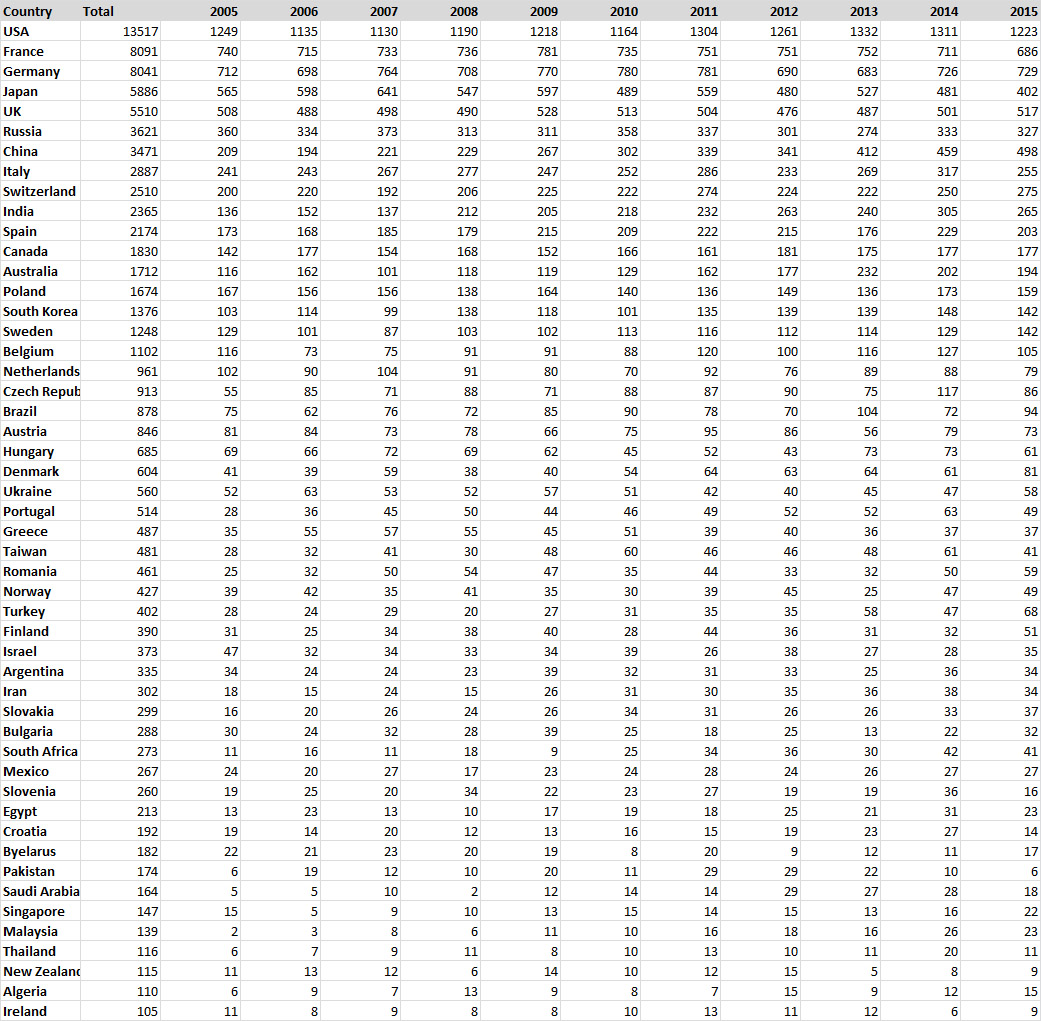}\\

\textbf{Table 2:} The number of publications for countries with more than 100 publications in the survey period.
\end{figure}

The number of publications in the survey period for countries in South and Middle America are stable ranging between 155 (2005) to 180 (2015) publications per year (Fig. \ref{Fig5}). Nearly 2/3 of the papers in total are from Brazil (51\%) and Argentina (19\%), followed by Mexico (15\%) (see Table 2).

\begin{figure}[tbp]
\centering
\includegraphics[width=0.4\linewidth]{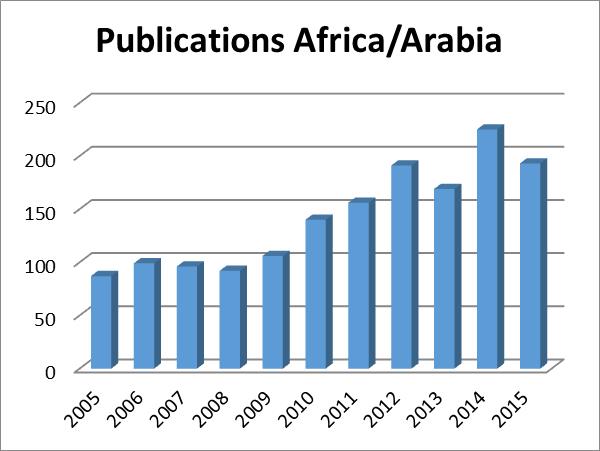}
\caption{Publications in Africa/Arabia per year in neutron scattering between 2005-2015.}
\label{Fig6}
\end{figure}

A strong increase also starting from a low basis is found in the number of publications for the countries in African and Arabic nations (Fig. \ref{Fig6}). Here the number has roughly doubled in the period from 2009 to 2015 (106 to 192). 

\subsection{Publications by towns were facilities located} Based on the town of the affiliation of the authors 23541 publications are related to a particular neutron facility in the time period of the survey. As the allocation is done only by the affiliation of the authors these numbers may differ from publication records published by individual facilities, but seem to express the productivity and use of the facilities within reasonable range. In particular the relative productivity and use can be expressed and also the quality of the papers related can be deduced. The numbers found show an average publication per year in a town hosting a facility of around 2100 papers in comparison to the total number of papers per year of about 4500. This may express that only every second paper may have an authorship or co-authorship of a facility member, which was used to relate a paper to a facility.

\begin{figure}[tbp]
\centering
\includegraphics[width=0.9\linewidth]{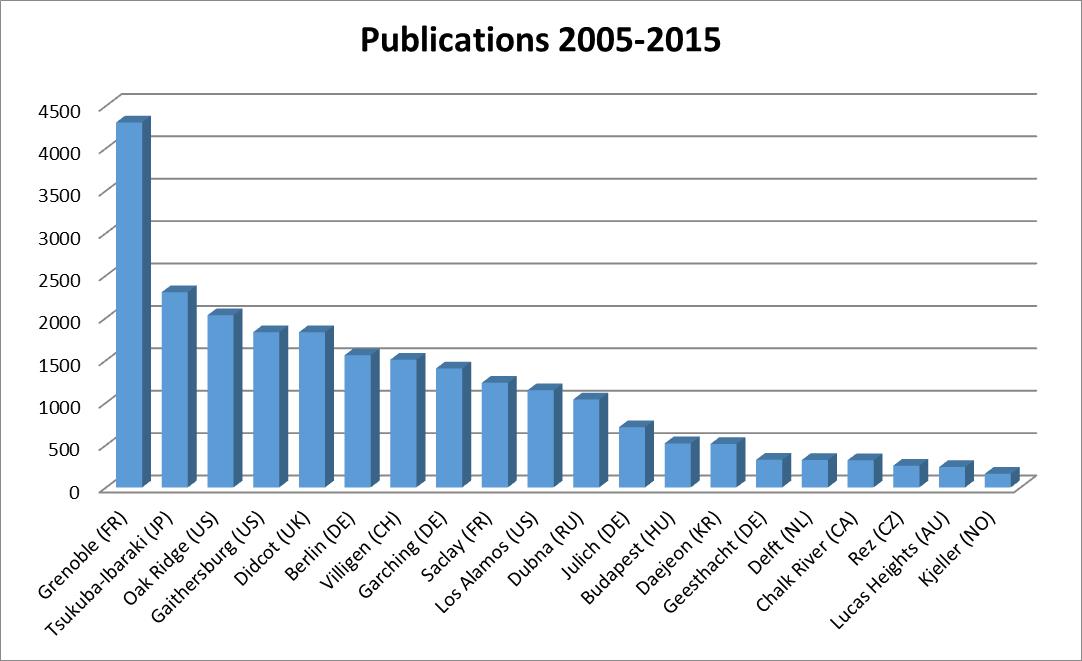}
\caption{Total number of publications by towns with neutron facilities in the period 2005-2015. (Note: Facilities in Geesthacht (DE) and J\"ulich (DE) were closed within the survey period and other facilities, such as SNS in Oak Ridge (US) or MLZ in Garching (DE) were ramped up.)}
\label{Fig7}
\end{figure}

According to the found numbers (Fig. \ref{Fig7}, Table 3), most productive facilities are the ILL (Grenoble, FR), J-PARC (Tsukuba-Ibaraki, JP), ORNL (SNS, HFIR) (Oak Ridge, US), NIST (Gaithersburg, US) and ISIS (Didcot, UK). The medium flux sources have a relative comparable productivity with BER II (Berlin, DE), SINQ (Villigen, CH), MLZ (Garching, DE), LLB (Saclay, FR), LANSCE (Los Alamos, US) and JINR (Dubna, RU). A clear increase is found for publications related to ORNL due to the increase in operation at SNS and HFIR in recent years. Also the MLZ shows a continuous increase in publications since start of operation in 2004. A decline is seen at the facilities BER II and LLB, both will be shut down end of 2019. 

\begin{figure*}[tbp]
\centering
\includegraphics[width=0.9\linewidth]{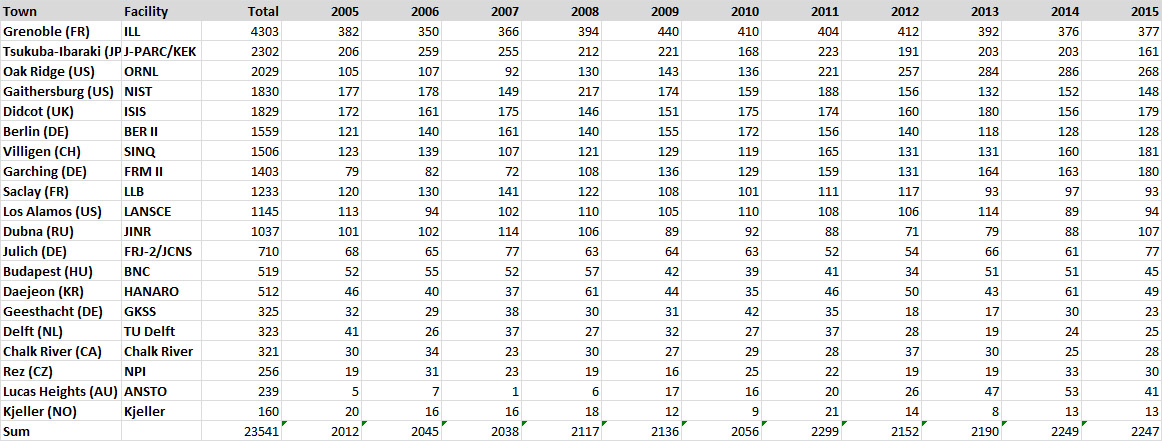}\\

\textbf{Table 3:} Number of publications by towns with neutron facilities in the period 2005-2015. (Note: Facilities in Geesthacht (DE) and J\"ulich (DE) were closed within the survey period, while others in Oak Ridge (US) or Garching (DE) were ramped up.)
\end{figure*}

\begin{figure}[tbp]
\centering
\includegraphics[width=0.9\linewidth]{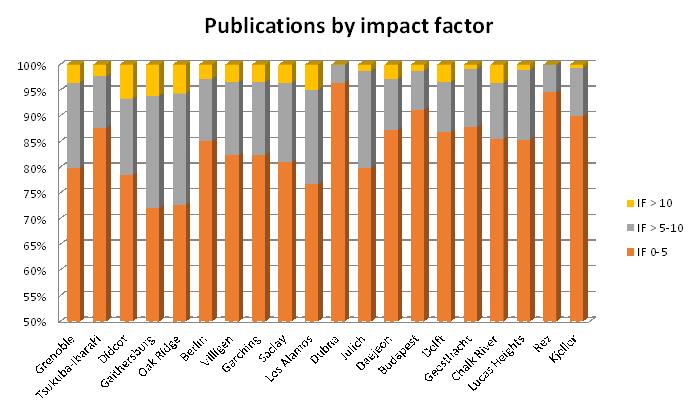}
\caption{Fraction of publications by impact factors and by towns with neutron facilities in the period 2005-2015.}
\label{Fig8}
\end{figure}

Analysing the publications according to the impact factor of the journal in which the paper has been published, some facilities as ILL, Grenoble, NIST, Gaithersburg, ORNL, Oak Ridge, and ISIS, Didcot, show a relative high fraction of publications in such journal (impact factor $>$5) (Fig. \ref{Fig8}). This is also reflected in high number of citations of publications related to these facilities. Here, in total 340251 citations were accumulated between 2005 to 2015 with 23541 publications related to facilities. On average these are 14,45 citations per publication. Analysing the number of citations of the publications related to a facility on average nearly 60\% of all publications have citations only up to 10 citations within the time period of the survey. On the other hand at some facilities up to 10\% of the publications reach citations above 50, e.g. at NIST, Gaithersburg.

\subsection{Publications by impact factor} 
\begin{figure}[tbp]
\centering
\includegraphics[width=0.5\linewidth]{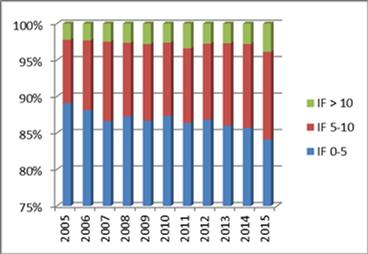}
\caption{Percentage of publications according to the journal impact factor.}
\label{Fig9}
\end{figure}

As a possible measure of the quality of the published papers within the survey, the impact factor of the journal in which the paper is published is often used as an indicator \cite{8,9}. Of the 49769 publications, 37761 are published in a journal with impact factor (IF) according to the "Web of Science" \cite{7}. More than 80\% of these papers are published in journals with IF up to 5 (Fig. \ref{Fig9}). This value is decreasing within the period of the survey from 89,1\% (2005) to 84,1\% (2015). In the same period publications in journals with IF between 5-10 and IF $>$10 have been increasing from 8,7\% to 12\% for IF 5-10 and 2,2\% to 3,9\% for IF $>$ 10. In particular the later increase by a factor of 1.7 an indicate a strong improvement in the quality and impact of publications using neutron scattering, also it cannot be earmarked in which field of science the scientific topics in the high impact journals where related to. It can, however, also reflect the fact that scientists are more and more pressured to publish in certain journals.

\begin{figure}[tbp]
\centering
\includegraphics[width=0.5\linewidth]{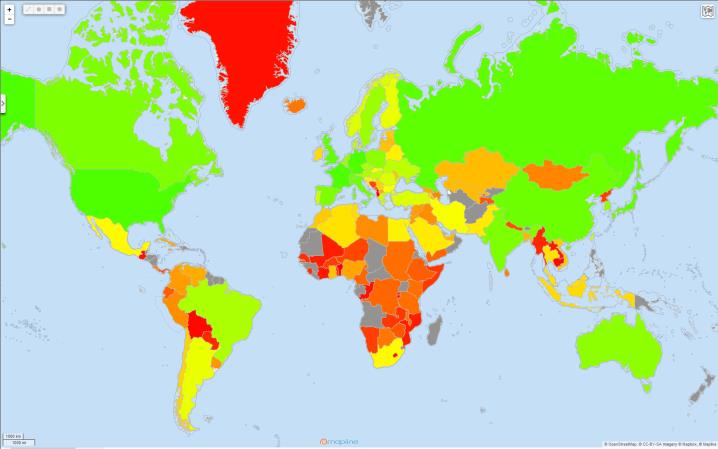}
\caption{Distribution of publications in journals with impact factor by world regions.}
\label{Fig10}
\end{figure}

Looking at the distribution of the publications with impact factors across world regions the majority is in Europe (52,6\%) followed by North America (18\%) and the Asia/Oceania (20,4\%) region. In North America more as 20\% of publications are in journals with impact factor of 5 or higher, 4,7\% have impact factors of above 10. For Europe the corresponding numbers are 15,7\% above 5 and 3,2\% above 10. Similar values are found for Asia/Ocania with 12,1\% above 5 and 2,5\% above 10. In all other regions only less than 5\% of papers are published in journals with impact factors above 5. This distribution reflects the higher productivity of science in the world regions due to the availability of neutron scattering facilities (Europe, North America, Asia) and the scientific infrastructure. In the map of the world below this distribution is expressed by colour code where green is related to the highest productivity and red lowest (Fig. \ref{Fig10}). Countries in grey do not appear in the publication record. 

In case of single facilities, the increase in publications in journals with higher impact factors is also reflected. E.g. at the MLZ the number of higher impact factor publications raised from 13\% (2005) to 25\% (2015), at the ILL it moved from 14\% (2005) to 21\% (2015).

In those countries with a stable and high publication rate, the relative number of publications in high impact factor journals is relatively stable (see e.g. USA or Germany). In raising countries the absolute as well as the relative numbers show a clear increase as seen e.g. in China, where the publication level in journals with IF $>$ 10 has raised from about 1\% in 2005 to above 5\% in 2015.

\subsection{Scientific topics} 
To get an insight into the scientific topics and fields most expressed in neutron scattering a survey according to keywords given in the abstracts of the publications counted was conducted. This exercise led to connectivity maps relating certain methods and areas, where scientific publications using neutron scattering are involved. The size of the circles around a keyword gives the quantity the keyword has been found. The lines and the strength of the lines express the connectivity of this keyword with other keywords.

\begin{figure}[bp]
\centering
\includegraphics[width=0.85\linewidth]{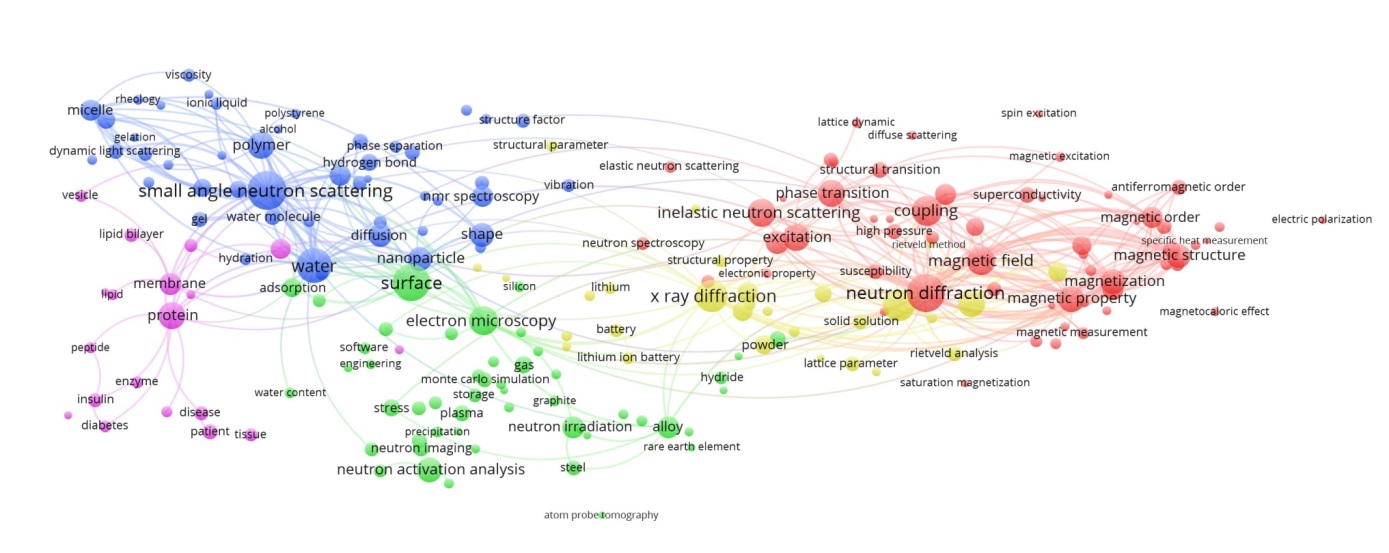}
\caption{Connectivity map of keywords in neutron scattering.}
\label{Fig11}
\end{figure}

In the overall connectivity of all keywords chosen distinct correlations between certain methods and scientific keywords are visible (Fig. \ref{Fig11}). E.g. the method neutron diffraction is close to keywords as magnetic structure while small angle neutron scattering is close to polymers or nanoparticles. Also complementary methods occur such as x-ray diffraction or electron microscopy.

Also these maps cannot be seen as a scientific description of a specific area and research field, they allow clustering the connectivity of certain keywords in the specific area towards each other and the occurrence with respect to its relevance in neutron scattering. In the example of the area of magnetism the keywords are closely connected while in material science a much broader and less dense connectivity with string subtopics is present.

\section{Conclusions}
\noindent
The present survey on the publication record using neutron scattering describes the publication record worldwide within the last ten years (2005-2015). On average neutron scattering publishes each year about 4500 papers which receive about 10000 citations, which demonstrates a strong and healthy scientific productivity of the neutron community. Within the period of the survey a distinct increase by emerging countries has happened. Countries like China or India have strong increases in their publication record as well as the whole Asia/Oceania region and the Africa/Arabic region. 

With an estimate of about 12000-14000 researchers using neutron scattering world-wide on average one publication per three researchers per year is produced. The capacity on neutron beam time is about 50000 beam days per year world-wide, estimated on numbers available for the European neutron facilities \cite{2}. This would relate to about 11 beam days for a single publication. The ILL Associates Report here estimates a number of 8 days, which reflects the higher availability of neutrons for science in Europe \cite{3}. The numbers also reflect a clear connection between capacity offered for neutrons and the publication record as demonstrated by the strong fraction of Europe on the total number of publications and citations.         

Regarding the quality of the publication record based on publications in journals with high and higher impact factors the relative number of papers in higher impact factors has increased strongly. Nowadays more as 15\% of publications are in journals with impact factor $>$5 compared to about 11\% in 2005.  In particular the publications related to the large facilities show a strong increase in high level publications. The ILL has increased the number of papers here from 14\% (2005) to 21\% (2015) and the MLZ, which started user operation just at the start of the survey period, from 13\% (2005) to 25\% (2015). The citation record of individual facilities shows an advantage of large facilities in comparison to medium or small facilities. E.g. at ILL or ISIS 37,1\% and 40,4\%, respectively, are cited more as 10 times, while at BER II or NPI only 29\% and 22,3\%, respectively, get this citation numbers.

Despite the questionable value of impact factors as indicators of scientific relevance and quality of research in combination with the citation record, the publication activities of the neutron community show a clear progress in total numbers of publications in particular in emerging regions and countries as Asia and China. Also the clear increase in publications in higher impact journals and the corresponding increase in citations are a sign of the healthy state of the community. Nevertheless the developed regions in Europe and the US will have to watch and deal with the ongoing developments. They will have to strengthen continuously their efforts in order to maintain their current leading role in this area of research to avoid loss in the world wide scientific and technological competitiveness \cite{2,10}.\\


\noindent\\
\textbf{\Large Appendix}

\noindent\\
\textbf{Requests to Web of Science}\\
Search criteria and strategy for publication survey on neutron scattering publications in the period 2005 – 2015 in the Web of Science. The following numbers of publications were obtained:

\noindent\\
1.	Basic request: 							64.800\\
2.	After selection of Web of Science Subject Categories:		51.000\\
3.	After selection of obtained journals:				50.800\\
4.	Final selection on the level of remaining journals:		49.769\\

\noindent
\textbf{1. Basic request:}\\
PY=(2005-2015) and TS=(Neutron$\$$scattering or Diffuse neutron scattering or Elastic neutron scattering or Incoherent neutron scattering or Inelastic neutron scattering or Magnetic neutron scattering or Neutron$\$$diffraction or Magnetic neutron diffraction or Neutron Laue diffraction or Neutron powder diffraction or Neutron powder$\$$diffraction or Neutron crystallography or Neutron protein crystallography or Neutron small angle scattering or Small angle neutron scattering or Small angle neutron$\$$scattering or Small$\$$angle neutron scattering or SANS or Neutron reflectometr* or Neutron reflectivit* or Neutron reflection or Neutron spectroscop* or Neutron inelastic spectroscop* or Inelastic neutron spectroscop* or INS or Quasi elastic neutron scattering or Quasi$\$$elastic neutron scattering or QENS or Neutron spin echo spectroscop* or Neutron$\$$spin echo or Neutron spin$\$$echo or Neutron resonance or Neutron spin resonance or Neutron spin$\$$resonance or Neutron three axis spectroscop* or Neutron three$\$$axis spectroscop* or Neutron triple$\$$axis spectroscop* or Triple$\$$axis spectroscop* or TAS or Neutron backscattering spectroscop* or Neutron radiograph* or Neutron tomograph* or Neutron imaging or Polari$\$$ed neutron* or Neutron polari$\$$ation or Neutron polari$\$$ation analys* or Neutron polarimetry* or Prompt gamma activation analys* or PGAA or Prompt gamma neutron activation analys* or Neutron activation analys* or Neutron depth profiling or Neutron interferometr* or Ultra$\$$cold neutron* or Ultracold neutron*)

\end{document}